\documentclass[12pt,preprint]{JHEP3}
\usepackage{graphicx}
\def\ie{{\it i.e.}}
\def\eg{{\it e.g.}}

\def\be{\begin{equation}}
\def\ee{\end{equation}}
\def\bea{\begin{eqnarray}}
\def\eea{\end{eqnarray}}
\def\bean{\begin{eqnarray*}}
\def\eean{\end{eqnarray*}}
\def\bary{\begin{array}}
\def\eary{\end{array}}

\def\bit{\begin{itemize}}
\def\eit{\end{itemize}}
\def\bwt{\begin{widetext}}
\def\ewt{\end{widetext}}

\def\lbar{\overline}
\def\nn{\nonumber}

\def\ub{{\bar u}}

\def\db{{\bar d}}
\def\sb{{\bar s}}
\def\bb{{\bar b}}

\title{Flavor SU(3) analysis of charmless $B$ meson decays to two pseudoscalar
  mesons}

\author{
Cheng-Wei Chiang\\
Department of Physics, National Central University,
Chungli, Taiwan 320, R.O.C. and\\
Institute of Physics, Academia Sinica, Taipei, Taiwan 115, R.O.C.\\
E-mail: \email{chengwei@phy.ncu.edu.tw}
}
\author{
Yu-Feng Zhou\\
Theory Group, KEK, Tsukuba, 305-0801, Japan.\\
E-mail: \email{zhou@post.kek.jp}
}

\date{\today}

\preprint{KEK TH-1103\\hep-ph/0609128}
\abstract{Global fits to charmless $B \to PP$ decays in the framework of flavor
  SU(3) symmetry are updated and improved without reference to the $\sin2\beta$
  measured from the charmonium decay modes.  Fit results directly constrain the
  $(\bar\rho,\bar\eta)$ vertex of the unitarity triangle, and are used to
  predict the branching ratios and CP asymmetries of all decay modes, including
  those of the $B_s$ system.  Different schemes of SU(3) breaking in decay
  amplitude sizes are analyzed.  The major breaking effect between
  strangeness-conserving and strangeness-changing decays can be accounted for
  by including a ratio of decay constants in tree and color-suppressed
  amplitudes.  The possibility of having a new physics contribution to $K \pi$
  decays is also examined from the data fitting point of view.}

\begin{document}

\section{Introduction \label{sec:intro}}

In the standard model (SM) of particle physics, CP violation in the quark
sector is postulated to be purely derived from the Kobayashi-Maskawa mechanism,
in which a $3 \times 3$ unitary CKM matrix $V_{\rm CKM}$ with three mixing
angles and one CP-violating phase is used to describe charge-current weak
transitions between the up-type and down-type quarks \cite{Cabibbo:1963yz}.
The unitarity condition on the first and third columns of the CKM matrix is
often used to form a triangle on the complex plane because the lengths of its
three sides are of the same order.  An important program in flavor physics is
to constrain this unitarity triangle using as many independent experimental
inputs as possible, for both determining standard model (SM) parameters with
high precisions and discovering any possible new physics effect.  A lot of
progress has been done in this direction with the help of a huge amount of $B$
meson data collected in the past few years at the $B$-factories
\cite{CKMfitter,UTfit}.

Although charmless modes are rare processes in $B$ decays, they are very
sensitive to the smallest CKM matrix elements $V_{ub}$ and $V_{td}$ through
decay amplitudes and mixing, respectively.  Moreover, they provide us
information about the weak phases associated with these two matrix elements.
Some theoretical analyses have been done in recent years to globally fit to
charmless $B \to PP$ and $VP$ decay data in the framework of QCD factorization
\cite{Beneke:2003zv} and flavor SU(3) symmetry
\cite{Chiang:2003pm,su3fit1,su3fit2,Malcles:2006jf}.  Here $P$ and $V$ refer to
pseudoscalar and vector mesons, respectively.  Since the weak phase $\beta$
($\phi_1$) is more accurately determined from the time-dependent CP asymmetry
analysis of $B_d \to ({\bar c}c) K_S$ modes, the result is usually used as an
input in the theory parameters.  With more modes being observed and measured at
higher precisions, it becomes possible to use purely rare decays to provide a
completely independent determination of the unitarity triangle vertex
$(\bar\rho,\bar\eta)$, expressed in terms of the Wolfenstein parameters
\cite{Wolfenstein:1983yz}, without reference to the charmonium modes.  It is
therefore one objective of the current analysis to see whether the charmless
$B$ decay data alone also provide a CKM picture consistent with other
constraints.

In this paper, we perform $\chi^2$ fits to the available charmless $B \to PP$
decays using the flavor diagram approach \cite{Zeppenfeld:1980ex}.  The fitting
parameters include the Wolfenstein parameters $A$, $\bar\rho$, and $\bar\eta$,
magnitudes of different flavor amplitudes, and their associated strong phases.
To take into account SU(3) breaking, we also include breaking factors of
amplitude sizes as our fitting parameters in some fits.  Generally speaking,
our fits render an area of the $(\bar\rho,\bar\eta)$ vertex slightly deviated
from but still consistent with that obtained from other constraints.  Aside
from a decay constant ratio, the SU(3) breaking is seen at $O(10)\%$ level.
From the extracted ranges of theory parameters, we predict the branching ratios
and CP asymmetries of all decays using flavor SU(3) symmetry, including the
$B_s$ system.  The latter will be compared with data already or to be measured
at the Tevatron, large hadron collider (LHC), and KEKB upgraded for running at
$\Upsilon(5S)$.

This paper is organized as follows.  In Section \ref{sec:notation}, we
introduce the notations used in this analysis, including the fitting
parameters.  Flavor amplitude decompositions of the rare decay modes, along
with the available experimental data on branching ratios and CP asymmetries,
are summarized in this section.  The fitting schemes and results are presented
in Section \ref{sec:fits}, where predictions and outlook of yet-seen modes are
also given.  Finally, Section \ref{sec:summary} summarizes our findings.

\section{Formalism and Notation \label{sec:notation}}

The magnitude of invariant decay amplitude $\cal A$ for a decay process $B \to
M_1 \, M_2$ is related to its partial width via the following relation:
\begin{eqnarray}
  \Gamma(B \to M_1 \, M_2)
  = \frac{|{\bf p}|}{8\pi m_B^2} |{\cal A}|^2 ~,
\end{eqnarray}
where $\bf p$ is the 3-momentum of the final state particles in the rest frame
of the $B$ meson.   To relate partial widths to branching ratios, we use the
world-average lifetimes $\tau^+ = (1.638 \pm 0.011)$ ps, $\tau^0 = (1.530 \pm
0.009)$ ps and $\tau_{B_s} = (1.466 \pm 0.059)$ ps computed by the Heavy Flavor
Averaging Group (HFAG) \cite{HFAG}.  Unless otherwise indicated, for each
branching ratio quoted we imply the average of a process and its $CP$-conjugate
one.

To perform the flavor amplitude decomposition, we use the following quark
content and phase conventions for mesons:
\begin{itemize}
\item{ {\it Bottom mesons}: $B^0=d\bb$, ${\lbar B}^0=b\db$, $B^+=u\bb$,
    $B^-=-b\ub$, $B_s=s\bb$, ${\lbar B}_s=b\sb$;}
\item{ {\it Pseudoscalar mesons}: $\pi^+=u\db$, $\pi^0=(d\db-u\ub)/\sqrt{2}$,
    $\pi^-=-d\ub$, $K^+=u\sb$, $K^0=d\sb$, ${\lbar K}^0=s\db$, $K^-=-s\ub$,
    $\eta=(s\sb-u\ub-d\db)/\sqrt{3}$,
    $\eta^{\prime}=(u\ub+d\db+2s\sb)/\sqrt{6}$;}
\end{itemize}
The $\eta$ and $\eta'$ mesons correspond to octet-singlet mixtures
\begin{eqnarray}
\eta  &=& \eta_8 \cos \theta_0 - \eta_1 \sin \theta_0 ~, \\
\eta' &=& \eta_8 \sin \theta_0 + \eta_1 \cos \theta_0 ~.
\end{eqnarray}
As shown in Ref.~\cite{Chiang:2003pm}, varying the mixing angle $\theta_0$ does
not improve the quality of fits.  Therefore, here we fix $\theta_0 =
\sin^{-1}(1/3) \simeq 19.5^\circ$ according to the above-mentioned quark
contents of $\eta$ and $\eta'$.

We list flavor amplitude decompositions and averaged experimental data for $B
\to PP$ decays in Tables \ref{tab:PPdS0} and \ref{tab:PPdS1}.  Values of
measured observables are obtained by weighted-averaging over the results of the
BaBar
\cite{Aubert:2005iy,Aubert:2005bq,Aubert:2006qd,Aubert:2006fy,Aubert:2006ad,%
  Aubert:2006ap,Aubert:2006fh,Aubert:2006gm,Lazzaro}, Belle
\cite{Chang:2004fz,Belle74,Abe:2006xp,Abe:2006xs,Abe:2006qx,Hara}, CLEO
\cite{Richichi:1999kj,Cronin-Hennessy:kg,Chen:2000hv,Behrens:1998dn,%
  Asner:2001eh}, and CDF \cite{Tonelli:2005cc,Maksimovic,Abulencia:2006ps}
Collaborations.  The standard deviation is scaled by the scale factor $S$
(whose definition can be found, for example, in Ref.~\cite{PDG}) if it is
greater than 1 in order to take into account the discrepancy among different
experimental groups.  These include: $Br(\pi^-\pi^0)$ ($S=1.3$),
$Br(\pi^-\eta)$ ($S=1.1$), $Br(\pi^-\eta')$ ($S=1.4$), ${\cal A}(\pi^+\pi^-)$
($S=2.6$), $Br(\pi^0\eta')$ ($S=1.4$), $A_{CP}(K^-\pi^0)$ ($S=1.3$), $Br(K^-
\eta)$ ($S=1.3$), $Br(\eta' {\bar K}^0)$ ($S=1.3$), and ${\cal A}(\eta' K_S)$
($S=1.4$).  Amplitudes such as annihilation and exchange diagrams are expected
to be small and therefore neglected in the calculation.

\TABLE{
\caption{Flavor amplitude decompositions for strangeness-conserving $B \to PP$
  decays.  Measured branching ratios (in units of $10^{-6}$) and CP asymmetries
  are given in the last two columns.  For those modes with time-dependent CP
  asymmetries, $\cal A$ and $\cal S$ are listed in the first and second rows,
  respectively.}
\label{tab:PPdS0}
\begin{tabular}{llccc}
\hline\hline
\multicolumn{2}{c}{Mode} & Flavor Amplitude
 & BR & ${\cal A}_{CP}$ \\ 
\hline
$B^- \to$
  & $\pi^- \pi^0$
    & $-\frac{1}{\sqrt{2}}(t+c)$
    & $5.7 \pm 0.5$
    & $0.04 \pm 0.05$ \\
& $K^- \lbar{K}^0$
    & $p$
    & $1.4 \pm 0.3$
    & $0.12 \pm 0.18$ \\
& $\pi^- \eta$
    & $-\frac{1}{\sqrt{3}}(t+c+2p+s)$
    & $4.4 \pm 0.4$
    & $-0.19 \pm 0.07$ \\
& $\pi^- \eta'$
    & $\frac{1}{\sqrt{6}}(t+c+2p+4s)$
    & $2.6 \pm 0.8$
    & $0.15 \pm 0.15$ \\
\hline
$\bar{B}^0 \to$
  & $K^+ K^-$
    & $-(e+pa)$ 
    & $0.07 \pm 0.11$
    & - \\
& $K^0\lbar{K}^0$
    & $p$ 
    & $1.0 \pm 0.2$
    & - \\
& $\pi^+ \pi^-$
    & $-(t+p)$
    & $5.2 \pm 0.2$
    & $0.39 \pm 0.19$ \\ 
& & & &  $-0.58 \pm 0.09$ \\
& $\pi^0 \pi^0$
    & $\frac{1}{\sqrt{2}}(-c+p)$ 
    & $1.3 \pm 0.2$
    & $0.36\pm0.32$ \\
& $\pi^0\eta$
    & $-\frac{1}{\sqrt{6}}(2p+s)$
    & $0.60\pm0.46$
    & - \\
& $\pi^0\eta'$
    & $\frac{1}{\sqrt{3}}(p+2s)$
    & $1.2 \pm 0.7$
    & - \\
& $\eta\eta$
    & $\frac{1}{3\sqrt{2}}(2c+2p+2s)$
    & $<1.2$
    & - \\
& $\eta\eta'$
    & $-\frac{1}{3\sqrt{2}}(2c+2p+5s)$
    & $<1.7$
    & - \\
& $\eta'\eta'$
    & $\frac{1}{3\sqrt{2}}(c+p+4s)$
    & $<10$
    & - \\
\hline
$\bar{B}_s^0 \to$
  & $K^+ \pi^-$
    & $-(t+p)$
    & $<5.6$
    & - \\
& $K^0 \pi^0$
    & $-\frac{1}{\sqrt{2}} (-c+p)$
    & -
    & - \\
& ${\bar K}^0 \eta$
    & $-\frac{1}{\sqrt{3}}(c+s)$
    & -
    & - \\
& ${\bar K}^0 \eta'$
    & $\frac{1}{\sqrt{6}}(c+3p+4s)$
    & -
    & - \\
\hline\hline
\end{tabular}
}

\TABLE{
\caption{Flavor amplitude decompositions for strangeness-changing $B \to PP$
  decays.  Measured branching ratios (in units of $10^{-6}$) and CP asymmetries
  are given in the last two columns.  For those modes with time-dependent CP
  asymmetries, $\cal A$ and $\cal S$ are listed in the first and second rows,
  respectively.}
\label{tab:PPdS1}
\begin{tabular}{llccc}
\hline\hline
\multicolumn{2}{c}{Mode} & Flavor Amplitude
 & BR & ${\cal A}_{CP}$ \\ 
\hline
$B^- \to$
  & $\pi^- \bar{K}^0$
    & $p'$ 
    & $23.1 \pm 1.0$
    & $0.01 \pm 0.02$ \\
& $\pi^0 K^-$
    & $-\frac{1}{\sqrt{2}}(p'+t'+c')$ 
    & $12.8 \pm 0.6$
    & $0.05 \pm 0.03$ \\
& $K^- \eta$
    & $-\frac{1}{\sqrt{3}}(s'+t'+c')$
    & $2.2 \pm 0.4$
    & $-0.29 \pm 0.11$ \\
& $K^- \eta'$
    & $\frac{1}{\sqrt{6}}(3p'+4s'+t'+c')$
    & $69.7 \pm 2.8$
    & $0.03 \pm 0.02$ \\
\hline
$\bar{B}^0 \to$
  & $\pi^+ K^-$
    & $-(p'+t')$
    & $19.7 \pm 0.6$
    & $-0.098 \pm 0.015$ \\
& $\pi^0 \bar{K}^0$
    & $\frac{1}{\sqrt{2}}(p'-c')$ 
    & $10.0 \pm 0.6$
    & $-0.12 \pm 0.11$ \\
& & & &  $0.33 \pm 0.21$ \\
& $\bar{K}^0 \eta$
    & $-\frac{1}{\sqrt{3}}(s'+c')$
    & $1.2\pm0.3$
    & - \\
& $\bar{K}^0 \eta'$
    & $\frac{1}{\sqrt{6}}(3p'+4s'+c')$
    & $64.9 \pm 4.4$
    & $-0.09 \pm 0.06$ \\
& & &  & $0.60 \pm 0.08$ \\
\hline
$\bar{B}_s^0 \to$
  & $K^+ K^-$
    & $-(p'+t')$
    & $34 \pm 9$
    & - \\
& $K^0 {\lbar K}^0$
    & $p'$
    & -
    & - \\
& $\pi^+ \pi^-$
    & $-(e'+pa')$
    & $<1.7$
    & - \\
& $\pi^0 \pi^0$
    & $\frac{1}{\sqrt{2}}(e'+pa')$
    & -
    & - \\
& $\pi^0 \eta$
    & $-\frac{1}{\sqrt{6}}c'$
    & -
    & - \\
& $\pi^0 \eta'$
    & $-\frac{1}{\sqrt{3}}c'$
    & -
    & - \\
& $\eta \eta$
    & $-\frac{1}{3\sqrt{2}}(2p'-2s'-2c')$
    & -
    & - \\
& $\eta \eta'$
    & $\frac{1}{3\sqrt{2}}(4p'+2s'-c')$
    & -
    & - \\
& $\eta' \eta'$
    & $\frac{1}{3\sqrt{2}}(4p'+8s'+2c')$
    & -
    & - \\
\hline\hline
\end{tabular}
}

In the present approximation, we consider five dominant types of independent
amplitudes: a ``tree'' contribution $T$; a ``color-suppressed'' contribution
$C$; a ``QCD penguin'' contribution $P$; a ``flavor-singlet'' contribution $S$,
and an ``electroweak (EW) penguin'' contribution $P_{EW}$.  The former four
types are considered as the leading-order amplitudes, while the last one is
higher order in weak interactions.  There are also other types of amplitudes,
such as the ``color-suppressed EW penguin'' diagram $P_{EW}^C$, ``exchange''
diagram $E$, ``annihilation'' diagram $A$, and ``penguin annihilation'' diagram
$PA$.  Due to dynamical suppressions, these amplitudes are ignored in the
analysis.  This agrees with the recent observation of the $B^0 \to K^+ K^-$
decay.

The QCD penguin amplitude in fact contains three components (apart from the CKM
factors): $P_t$, $P_c$, and $P_u$, with the subscript denoting which quark is
running in the loop.  After imposing the unitarity condition, we are left with
two components: $P_{tc} = P_t - P_c$ and $P_{tu} = P_t - P_u$, integrating out
the $t$ quark from the theory.  For simplicity, we will assume the $t$-penguin
dominance, so that $P_{tc} = P_{tu}$ and are denoted by a single symbol $P$.
The same comment applies to other penguin-type amplitudes (\eg, $P_{EW}$ and
$P_{EW}^C$) as well.

In physical processes, the above-mentioned flavor amplitudes always appear in
specific combinations.  To simplify the notations, we therefore define the
following unprimed and primed symbols for $\Delta S = 0$ and $|\Delta S| = 1$
transitions, respectively:
\begin{eqnarray}
\label{eqn:dict}
t \equiv Y_{db}^u T - (Y_{db}^u + Y_{db}^c) P_{EW}^C ~,
&\quad&
t' \equiv Y_{sb}^u \xi_t T - (Y_{sb}^u + Y_{sb}^c) P_{EW}^C ~, \nn \\
c \equiv Y_{db}^u C - (Y_{db}^u + Y_{db}^c) P_{EW} ~,
&\quad&
c' \equiv Y_{sb}^u \xi_c C - (Y_{sb}^u + Y_{sb}^c) P_{EW} ~, \nn \\
p \equiv - (Y_{db}^u + Y_{db}^c) \left( P - \frac{1}{3} P_{EW}^C \right) ~,
&\quad&
p' \equiv - (Y_{sb}^u + Y_{sb}^c) \left( \xi_p P - \frac{1}{3} P_{EW}^C \right)
~,
\\
s \equiv - (Y_{db}^u + Y_{db}^c) \left( S - \frac{1}{3} P_{EW} \right) ~,
&\quad&
s' \equiv - (Y_{sb}^u + Y_{sb}^c) \left( \xi_s S - \frac{1}{3} P_{EW} \right)
~,
\nn
\\
a \equiv Y_{db}^u A ~,
&\quad&
a' \equiv Y_{sb}^u A ~, \nn \\
e \equiv Y_{db}^u E - (Y_{db}^u + Y_{db}^c) PA ~,
&\quad&
e' \equiv Y_{sb}^u E - (Y_{sb}^u + Y_{sb}^c) PA ~, \nn
\end{eqnarray}
where $Y_{qb}^{q'} \equiv V_{q'q} V_{q'b}^*$.  Unless they are leading
contributions, amplitudes such as $e$ and $pa$ are omitted from Tables
\ref{tab:PPdS0} and \ref{tab:PPdS1}.

One differnce between the current analysis and our previous work
\cite{Chiang:2003pm} is that the CKM matrix elements associated with the
amplitudes are factored out here.  The strong phases, however, are still
absorbed in the amplitudes.  Notice that when going from $\Delta S = 0$ to
$|\Delta S| = 1$ transitions, we explicitly include SU(3) breaking factors
$\xi_t, \xi_c$, and $\xi_p$ for the $T$, $C$, and $P$ amplitudes, respectively.
In the naive factorization approximation, these SU(3) breaking factors are all
equal to $\xi \equiv f_K/f_{\pi} = 1.223$ \cite{PDG}.  As an example, using the
above-defined notations we have for the $B^0 \to K^+\pi^-$ decay:
\begin{eqnarray*}
  \mathcal{A}(K^+\pi^-) &=&
  - Y_{sb}^{u} \xi_t T + \left( Y_{sb}^{u} + Y_{sb}^{c} \right) \xi_p P ~.
\end{eqnarray*}
This can be obtained from the complete set of flavor amplitude decompositions
given in Table \ref{tab:PPdS1}.  

The CKM factors used in the analysis are given in terms of the Wolfenstein
parameterization of the CKM matrix to $O(\lambda^5)$.  Since $\lambda$ has been
determined from kaon decays to a high accuracy, we will simply use the central
value $0.2272$ quoted by the CKMfitter group \cite{CKMfitter} as a theory
input, and leave $A$, $\bar\rho \equiv \rho ( 1 - \lambda^2/2 )$, and $\bar\eta
\equiv \eta ( 1 - \lambda^2/2 )$ as fitting parameters to be determined by data.

A relation between the sizes of EW penguin amplitude and tree-type amplitudes
has been found in Ref.~\cite{NR} where the Fierz transformation is used to
relate EW penguin operators with tree-level operators.  Explicitly,
\begin{eqnarray}
  P_{EW} = - \delta_{EW} |T + C| e^{i \delta_{P_{EW}} }~,
\end{eqnarray}
where $\delta_{P_{EW}}$ is the strong phase associated with $P_{EW}$.  In the SM,
\begin{eqnarray}
\label{eq:dEW}
\delta_{EW} \simeq
-\frac{3}{2}\frac{C_{9}+C_{10}}{C_{1}+C_{2}} = 0.0135 \pm 0.0012 ~.
\end{eqnarray}
In our fit, we will leave it as a free parameter to test how well the above
relation holds.

For the $B$ meson decaying into a CP eigenstate $f_{CP}$, the time-dependent CP
asymmetry is written as
\begin{eqnarray}
\label{eq:t-dep}
A_{CP}(t)
&=& 
\frac{\Gamma(\bar{B}^{0}\to f_{CP})-\Gamma(B^{0}\to f_{CP})}
     {\Gamma(\bar{B}^{0}\to f_{CP})+\Gamma(B^{0}\to f_{CP})} \nn \\
&=&
{\cal S} \, \sin(\Delta m_{B} \cdot t) 
+ {\cal A} \, \cos(\Delta m_{B} \cdot t) ~,
\end{eqnarray}
where $\Delta m_{B}$ is the mass difference between the two mass eigenstates of
$B$ mesons and $t$ is the decay time measured from the tagged $B$ meson.

\section{Fitting Analysis \label{sec:fits}}

To see the effects of SU(3) symmetry breaking, we consider the following four
fitting schemes in our analysis:
\begin{enumerate}
\item exact flavor SU(3) symmetry for all amplitudes (\ie, $\xi_t = \xi_c =
  \xi_p = 1$);
\item including the factor $f_K/f_\pi$ for $|T|$ only (\ie, $\xi_t = f_K/f_\pi$
  and $\xi_c = \xi_p = 1$);
\item including the factor $f_K/f_\pi$ for both $|T|$ and $|C|$ (\ie, $\xi_t =
  \xi_c = f_K/f_\pi$ and $\xi_p = 1$); and
\item including a universal SU(3) breaking factor $\xi$ for all amplitudes on
  top of Scheme~3 (\ie, $\xi_t = \xi_c = \xi f_K/f_\pi$ and $\xi_p = \xi$).
\end{enumerate}
To reduce the number of parameters, we assume exact flavor SU(3) symmetry for
the strong phases in these fits. As a phase convention we set the tree
amplitude to be real and positive, \ie, $\delta_T=0$.

In addition to the observables in $B \to PP$ modes, we also include $|V_{ub}| =
(4.26 \pm 0.36) \times 10^{-3}$ and $|V_{cb}| = (41.63 \pm 0.65) \times
10^{-3}$ as our fitting observables.  Here we take the averages of their values
measured from inclusive and exclusive decays as quoted in
Ref.~\cite{CKMfitter}.  They mainly help fixing the values of $A$ and
$\sqrt{\rho^2 + \eta^2}$.  We will discuss below how our fit results and
predictions change should we use a lower value of $V_{ub}$ in the following
numerical analysis.

\subsection{Fits to modes with only $\pi, K$ mesons in the final state
  \label{sec:fitsome}}

We start by fitting to the branching ratios and CP asymmetries of the $\pi\pi$,
$\pi K$, and $KK$ modes of $B$ meson decays.  This part of the analysis avoid
uncertainties in the wave functions and singlet amplitudes associated with the
$\eta$ and $\eta'$ mesons.

Currently, there are 20 experimental observables.  Along with $|V_{ub}|$ and
$|V_{cb}|$ mentiones above, we have totally 22 data points.  The number of
theoretical parameters is 10 for Schemes~1 to 3 and 11 for Scheme~4.  The
best-fitted values of the parameters in their 1 $\sigma$ ranges along with the
minimal $\chi^2$ values, $\chi^2_{\rm min}$, for the different schemes are
listed in Table \ref{tab:parafit}.

\TABLE{
\caption{Fit results of the parameters for the $\pi\pi$, $\pi K$, and $KK$
  modes in Schemes 1 through 4 defined in the text along with the minimal
  $\chi^2$ value.  The amplitudes are given in units of $10^4$ eV.}
\label{tab:parafit}
\begin{tabular}{ccccc}
\hline\hline
Parameter & Scheme 1 & Scheme 2 & Scheme 3 & Scheme 4 \\
\hline
$\bar{\rho}$                 &$0.139^{+0.042}_{-0.037}$&$0.134^{+0.041}_{-0.036}$&$0.134^{+0.041}_{-0.036}$&$0.133^{+0.039}_{-0.035}$\\
$\bar{\eta}$                 &$0.401\pm0.030$          &$0.403\pm0.031$          &$0.404\pm0.031$          &$0.399\pm0.031$          \\
$A$                     &$0.807\pm0.013$          &$0.807\pm0.013$          &$0.807\pm0.013$          &$0.807\pm0.013$          \\
\hline
$|T|$                        &$0.573^{+0.055}_{-0.047}$&$0.575^{+0.055}_{-0.047}$&$0.574^{+0.055}_{-0.047}$&$0.582^{+0.056}_{-0.049}$\\
$|C|$                        &$0.371\pm0.050$          &$0.364\pm0.050$          &$0.364\pm0.049$          &$0.372\pm0.051$          \\
$\delta_C$                   &$-57.6\pm10.3$           &$-55.9\pm10.7$           &$-55.8\pm10.2$           &$-56.3\pm10.1$           \\
$|P|$                        &$0.121\pm0.002$          &$0.122\pm0.002$          &$0.122\pm0.002$          &$0.117\pm0.008$          \\
$\delta_P$                   &$-22.7\pm4.0$            &$-18.8\pm3.2$            &$-19.3\pm3.2$            &$-18.6^{+3.2}_{-3.5}$    \\
$|P_{EW}|$                      &$0.011^{+0.006}_{-0.003}$&$0.011^{+0.006}_{-0.003}$&$0.011^{+0.005}_{-0.003}$&$0.011^{+0.004}_{-0.003}$\\
$\delta_{P_{EW}}$                 &$-4.3^{+34.1}_{-50.6}$   &$2.2^{+32.0}_{-49.3}$    &$-10.0^{+37.2}_{-45.3}$  &$-15.1\pm39.9$           \\
$\xi$ &1(fixed)&1(fixed)&1(fixed)&$1.04^{+0.08}_{-0.07}$\\
\hline
$\delta_{EW}$ &$0.013\pm0.006$&$0.013\pm0.006$&$0.013\pm0.005$&$0.013\pm0.004$\\
\hline
$\chi^2_{\rm min}/dof$                    &18.9/12                     &18.0/12                     &16.4/12                     &16.1/11                     \\
\hline\hline
\end{tabular}
}

Generally speaking, we obtain fairly stable results for the parameters, except
for some small variations in the strong phases among the SU(3) breaking schemes
considered here.  The fit quality is best in Scheme~3, suggesting that it is
better to include the SU(3) breaking factor $f_K/f_\pi$ for the $T$ and $C$
amplitudes when going from the strangeness-conserving modes to the
strangeness-changing modes.  Scheme~4 introduces an additional SU(3) breaking
factor $\xi$, which is found to be about 1.04.  The small difference in
$\chi^2_{\rm min}$ between Scheme~3 and Scheme~4 turns out to reduce the
fitting quality from 17\% down to 14\%.

It has been suggested as a direct test of flavor SU(3) symmetry by comparing
the extracted amplitude magnitudes of $B^0 \to K^0 \overline{K}^0$ and $B^+ \to
K^+ \overline{K}^0$ with $B^+ \to K^0 \pi^+$ because all of them have the same
single penguin amplitude contributing to the decays, except for the only
difference in the CKM factors.  This is verified experimentally according to
the current data.  Taking the averaged invariant amplitudes of $\bar{B}^0 \to
K^0 \overline{K}^0$ and $B^- \to K^- \overline{K}^0$ as $|p|$ and comparing it
with $|p'|$ obtained from $B^+ \to K^0 \pi^+$, one obtains $|p/p'| \simeq 0.23
\pm 0.02$ consistent with the expected ratio $|V_{cd}/V_{cs}|$.  This justifies
our choice of not including the factor $f_K / f_\pi$ for SU(3) breaking in the
penguin amplitudes.  We also find that $\chi^2_{\rm min}$ becomes worse when
the factor is imposed on the QCD penguin amplitude.  Therefore, the data
indicate that to a good approximation factorization works better for the
color-allowed and color-suppressed tree amplitudes (\ie, $T$ and $C$).

We observe a large $|C|/|T|$ ratio of about $0.63$ in all these fits.  This is
different from the expectation of the usual perturbative calculation within the
SM.  A possible explanation is given in Ref.~\cite{Li:2005kt} where
next-to-leading order corrections to the interaction vertex are found to
enhance the color-suppressed amplitude $C$ for $K \pi$ decays by a factor of 2
to 3 while keeping other amplitudes more or less unchanged.  Moreover, there
exists a large relative strong phase of about $(-56 \pm 10)^\circ$ between $C$
and $T$.  Therefore, it can play an important role in CP asymmetries.  The
values of these parameters are largely driven by the large branching ratio of
$\pi^0 \pi^0$ and $A_{CP}(\pi^0 K^-)$ being quite different from
$A_{CP}(\pi^+K^-)$.

A strong phase of about $-20^\circ$ is associated with the penguin amplitude.
This is primarily demanded by the CP asymmetries of the $\pi^+ \pi^-$ and $K^+
\pi^-$ modes, both of which involve the combination of $t^{(\prime)}$ and
$p^{(\prime)}$.

In all our fits, the parameter $\delta_{EW}$ is seen to be very stable and
close to the value in Eq.~(\ref{eq:dEW}), which shows that the EW penguin
amplitude has a size roughly agreeing with the SM expectation. This is partly
due to the fact that the latest data are moving towards the SM estimates and a
larger best fitted $\gamma$ is favored, modifying the $T-P$ and $T-P_{EW}$
interferences and enhancing  $Br(\pi^0\bar{K}^0)$.  It does not necessarily
mean that the possibility of new physics has been ruled out. As it will be
shown below (see Section \ref{sec:fitNP}), an electroweak penguin-like new
contribution with a different CP phase can dramatically improve the quality of
fits.  Besides, we find that $P_{EW}$ has a strong phase of about $-10^\circ$
relative to $T$ and about $45^\circ$ to $C$.

\FIGURE{
\centering
\includegraphics[width=10cm]{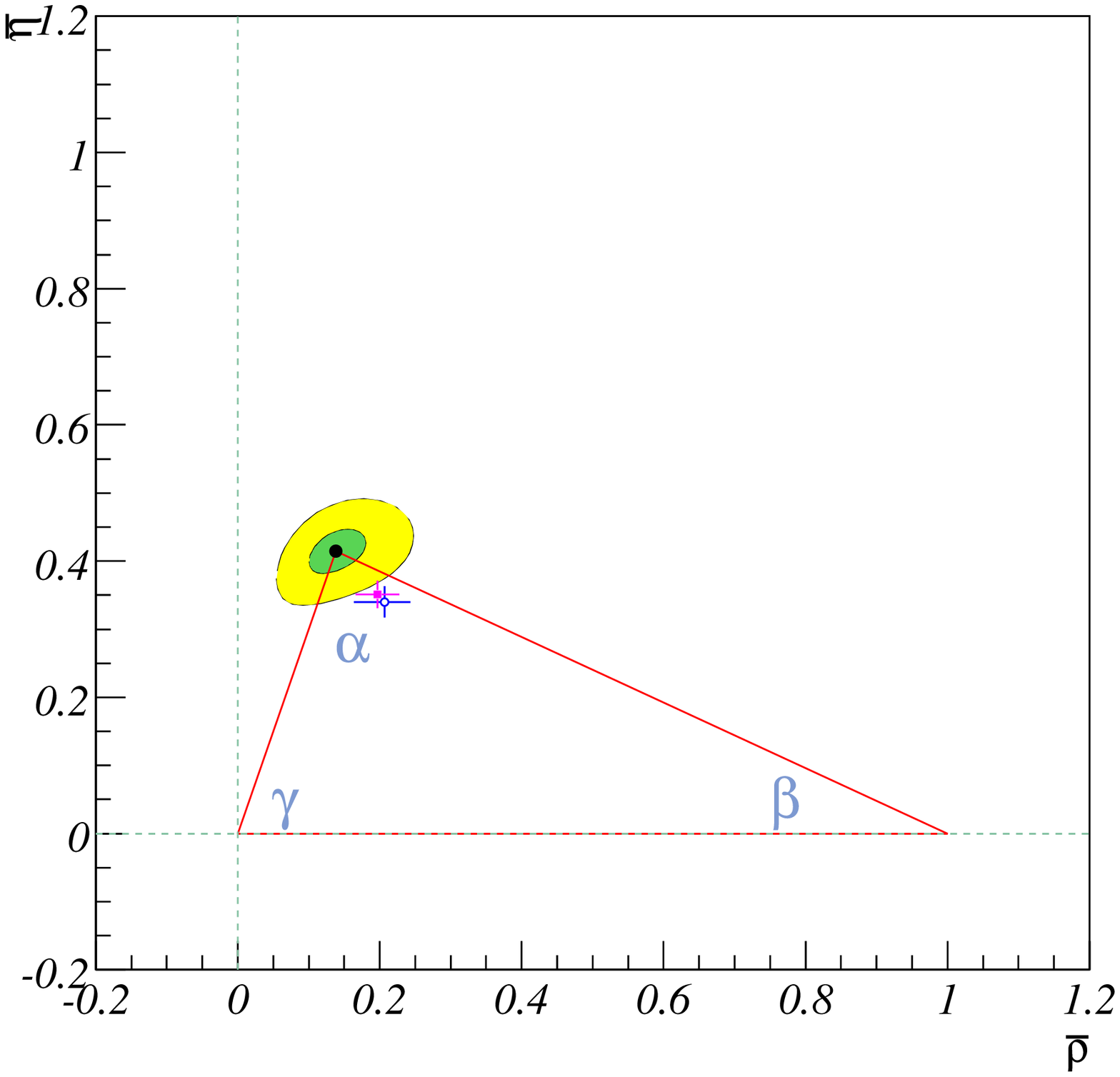}
\caption{Constraints on the $(\bar\rho,\bar\eta)$ vertex using $B \to \pi \pi,
  K \pi$, and $KK$ data in Scheme~3 defined in the text.  Contours correspond
  to 1 $\sigma$ and $95\%$ CL, respectively.  The crosses refer to the 1
  $\sigma$ range given by the latest CKMfitter (open circle) and UTfit (filled
  square) results using other methods \cite{CKMfitter,UTfit} as a comparison.}
\label{fig:rhoeta}
}

Since the 1 $\sigma$ and 95\%CL ranges for the $(\bar\rho,\bar\eta)$ vertex
have unnoticeable change in the four fitting schemes defined above, we only
present as a representative our preferred set, Scheme~3, in
Fig.~\ref{fig:rhoeta}.  This also shows the stability of the
$(\bar\rho,\bar\eta)$ values against SU(3) breaking.  Scheme~3 gives the
following results for the weak phases $\alpha$, $\beta$, and $\gamma$:
\begin{eqnarray}
  &
  \alpha = \left( 83^{+6}_{-7} \right)^\circ ~, \quad\mbox{or}\quad
  69^\circ \le \alpha \le 96^\circ \quad \mbox{(95\% CL)} ~;
  &
  \nonumber \\
  &
  \beta = \left( 26 \pm 2 \right)^\circ ~, \quad\mbox{or}\quad
  21^\circ \le \beta \le 31^\circ \quad \mbox{(95\% CL)} ~;
  &
  \\
  &
  \gamma = \left( 72^{+4}_{-5} \right)^\circ ~, \quad\mbox{or}\quad
  62^\circ \le \gamma \le 81^\circ \quad \mbox{(95\% CL)} ~.
  &
  \nonumber
\end{eqnarray}

As shown in Fig.~\ref{fig:rhoeta}, the determined ranges of
$(\bar\rho,\bar\eta)$ for all the schemes in our fitting are slightly higher
than those given by the latest CKMfitter and UTfit results obtained using other
observables \cite{CKMfitter,UTfit}:
\begin{eqnarray}
\mbox{CKMfitter:} &&
\bar\rho = 0.207^{+0.036}_{-0.034} ~, \; \bar\eta = 0.341 \pm 0.023 ~;
\nonumber \\
\mbox{UTfit:} &&
\bar\rho = 0.197 \pm 0.031 ~, \; \bar\eta = 0.351 \pm 0.020 ~.
\end{eqnarray}
These values are indicated by crosses with an open circle and a filled square
in the figure, respectively.  The difference is to a large extent caused by the
large value of $|V_{ub}|$ used in our fits.  Therefore, we obtain slightly
larger phases $\beta$ and $\gamma$ but a smaller $\alpha$.

We now briefly comment on the effects of using a smaller value of $|V_{ub}|$ in
the fits.  If we take $|V_{ub}| = (3.50 \pm 0.18) \times 10^{-3}$ extracted
from unitarity angle measurements only \cite{Bona:2006ah}, $\chi^2_{\rm min}$
is improved by $1.1$.  $\beta$ reduces to around $21^\circ$ and $\gamma$
increases to about $75^\circ$, with $\alpha$ almost unaffected.  The magnitudes
of $|T|$ and $|C|$ both become slightly larger, but their ratio stays the same.
The other parameters do not change much either.  The same features are also
observed in fits with all $PP$ modes to be discussed in the next section.
However, it should be emphasized that our analysis purposely avoid inputs other
than the charmless decay modes unless necessary (such as $\lambda$ and
$|V_{cb}|$ mentioned above).  We therefore do not use this smaller $|V_{ub}|$
value in our main analysis, for it relies quite a lot on charmed $B$ decays.

Our predictions for the branching ratios and CP asymmetries for all the
$B_{u,d} \to \pi \pi, K \pi$, and $KK$ modes based upon the extracted
parameters in Table~\ref{tab:parafit} are given in Table~\ref{tab:br-acp}.

\TABLE{
\caption{Predicted branching ratios (in units of $10^{-6}$) and CP asymmetries
  for $B \to \pi\pi, K\pi$, and $K K$ modes in different schemes.  The
  observables with vanishing entries are predicted to be identically zero in
  our analysis.  Experimentally measured quantities, if any, are already given
  in the last two columns of Tables~\ref{tab:PPdS0} and \ref{tab:PPdS1} for
  comparison.}
\label{tab:br-acp}
\begin{tabular}{ccccc}
\hline\hline
Observable & Scheme 1 & Scheme 2 & Scheme 3 & Scheme 4  \\
\hline
$Br(\pi^+\pi^-)$         &$5.4\pm1.1$              &$5.4\pm1.0$              &$5.3\pm1.0$              &$5.3\pm1.1$              \\
$Br(\pi^0\pi^0)$         &$1.6\pm0.4$              &$1.6\pm0.4$              &$1.6\pm0.4$              &$1.5\pm0.4$              \\
$Br(\pi^-\pi^0)$         &$5.3\pm1.2$              &$5.4\pm1.2$              &$5.4\pm1.2$              &$5.4\pm1.3$              \\
$Br(\pi^+K^-)$           &$20.2\pm1.0$             &$20.1\pm1.1$             &$20.1\pm1.1$             &$20.3\pm4.3$             \\
$Br(\pi^0\bar{K}^0)$     &$9.9\pm1.0$              &$9.9\pm1.0$              &$10.0\pm0.9$             &$10.1\pm2.3$             \\
$Br(\pi^-\bar{K}^0)$     &$23.0\pm1.1$             &$23.1\pm1.1$             &$23.1\pm1.1$             &$23.4\pm4.8$             \\
$Br(\pi^0K^-)$           &$12.0\pm1.2$             &$12.1\pm1.2$             &$12.0\pm1.1$             &$12.2\pm2.5$             \\
$Br(K^+K^-)$             &$0$              &$0$              &$0$              &$0$              \\
$Br(K^0\bar{K}^0)$       &$1.0\pm0.1$              &$1.0\pm0.1$              &$1.0\pm0.1$              &$1.0\pm0.2$              \\
$Br(K^-\bar{K}^0)$       &$1.1\pm0.1$              &$1.1\pm0.1$              &$1.1\pm0.1$              &$1.0\pm0.2$              \\
${\cal A}(\pi^+\pi^-)$         &$0.32\pm0.07$            &$0.27\pm0.06$            &$0.28\pm0.06$            &$0.26\pm0.06$            \\
${\cal A}(\pi^0\pi^0)$         &$0.47\pm0.15$            &$0.49\pm0.15$            &$0.49\pm0.14$            &$0.50\pm0.14$            \\
$A_{CP}(\pi^-\pi^0)$         &$-0.01\pm0.04$           &$-0.02\pm0.03$           &$-0.01\pm0.03$           &$-0.01\pm0.03$           \\
$A_{CP}(\pi^+K^-)$           &$-0.08\pm0.02$           &$-0.09\pm0.02$           &$-0.09\pm0.02$           &$-0.09\pm0.02$           \\
${\cal A}(\pi^0 K_S)$     &$-0.07\pm0.03$           &$-0.08\pm0.02$           &$-0.09\pm0.03$           &$-0.10\pm0.03$           \\
$A_{CP}(\pi^-\bar{K}^0)$     &$0$            &$0$            &$0$            &$0$            \\
$A_{CP}(\pi^0K^-)$           &$0.00\pm0.03$            &$0.00\pm0.03$            &$0.01\pm0.04$            &$0.02\pm0.04$            \\
$A_{CP}(K^+K^-)$             &$0$            &$0$            &$0$            &$0$            \\
${\cal A}(K^0\bar{K}^0)$       &$0$            &$0$            &$0$            &$0$            \\
$A_{CP}(K^-\bar{K}^0)$       &$0$            &$0$            &$0$            &$0$            \\
${\cal S}(\pi^+\pi^-)$          &$-0.580\pm0.130$         &$-0.585\pm0.130$         &$-0.584\pm0.130$         &$-0.565\pm0.141$         \\
${\cal S}(\pi^0\pi^0)$          &$0.814\pm0.109$          &$0.812\pm0.108$          &$0.810\pm0.106$          &$0.786\pm0.113$          \\
${\cal S}(\pi^0 K_S)$      &$0.851\pm0.042$          &$0.850\pm0.041$
&$0.861\pm0.041$         &$0.858\pm0.042$         \\
${\cal S}(K^0\bar{K}^0)$        &$-0.000\pm0.014$         &$-0.000\pm0.014$         &$-0.000\pm0.014$         &$-0.000\pm0.015$         \\
\hline\hline
\end{tabular}
}

It is seen that the values in the table are quite stable and generally in
agreement with the measured numbers or upper bounds within the errors.  The
largest $\chi^2$ comes from ${\cal S}(\pi^0 K_S)$, which is entailed to be even
larger than $S_{(c{\bar c})K_S}$.  The CP asymmetry of $K^- \pi^0$ is found to
be close to zero, giving the second largest contribution to $\chi^2$.  As we
will see in Section~\ref{sec:fitNP}, these discrepancies can be significantly
reduced if a new amplitude is introduced.

The central values of ${\cal A}(\pi^+\pi^-)$ are slightly smaller than the
measured one, but deviate from zero at more than 4 $\sigma$ level.  The
predicted ${\cal A}(\pi^0 \pi^0)$ are noticeably different from zero.  This is
seen as a result of the absence of a dominant amplitude in the decay.  We also
find a sizeable ${\cal S}(\pi^0 \pi^0) \sim 0.8 \pm 0.1$, to be verified
experimentally.

At this point, it may be helpful to compare the predictions of other
approaches.  The recent next-to-leading order (NLO) calculations in QCD
factorization (QCDF) show some enhancements from strong penguin corrections at
${\cal O}(\alpha_s^2)$ \cite{Li:2005wx}.  However, their predictions for
CP-averaged decay rates of the $\pi K$ modes still tend to be lower than the
experimental data using the default parameter set, and the observed large
negative $A_{CP}(\pi^+K^-)$ is difficult to understand.  The hard
spectator-scattering corrections have also been calculated to this order, and
are found to possibly have a significant impact on the tree-dominated $B \to
\pi \pi$ decays \cite{Beneke:2005vv}.  It remains to be seen if a complete
next-to-next-to-leading order (NNLO) calculation can lead to a better agreement
with the data.

In the perturbative QCD (pQCD) approach, NLO calculations including vertex
corrections, quark loops and magnetic penguins suppress $A_{CP}(\pi^0K^-)$
while keep $A_{CP}(\pi^+K^-)$ large enough with the correct sign
\cite{Li:2005kt}, both in good agreement with the data.  Nevertheless, the
corrections to $\pi\pi$ modes are ineffective so that the predicted
$Br(\pi^0\pi^0)$ remains small.

A recent comprehensive analysis in the same set of observables has been
performed in the heavy quark limit of QCD and in the soft-collinear effective
theory (SCET) \cite{Bauer:2005kd}.  This approach generally involves more
hadronic parameters without the help of symmetries.  At the LO, they predict a
larger $Br(\pi^+K^-)$, an $A_{CP}(\pi^0 K^-)$ close to $A_{CP}(\pi^+K^-)$, and
an ${\cal A} (\pi^0 {\bar K}^0)$ opposite in sign to the data.

In all these approaches, the predicted $S(\pi^0K_S)$ are close to the one from
$S(J/\psi K_S)$ or even larger \cite{Beneke:2005pu,Li:2005kt,Bauer:2005kd}.
This leaves room for new physics interpretations if it is further confirmed by
data at a higher precision.

Our predictions for the branching ratios and CP asymmetries for all the $B_s
\to \pi \pi, K \pi$, and $KK$ modes based upon the extracted parameters in
Table~\ref{tab:parafit} are given in Table~\ref{tab:predBs}.  They serve as
another good testing ground for the flavor SU(3) symmetry.  Among all
observables of the $B_s$ decays only the branching ratio of the $B_s \to K^+
K^-$ mode is observed at CDF \cite{Tonelli:2005cc}.  This mode has the same
flavor amplitude decomposition as the $\bar{B}^0 \to \pi^+ K^-$ mode.
Therefore, our predictions in this case are close to those for $\bar{B}^0 \to
\pi^+ K^-$, apart from a small difference due to such factors as the masses and
decay widths.  They are all smaller than the measured value.  Since this is
only observed for the first time with somewhat large errors, a more precise
determination will be very helpful.  Besides, this mode is predicted to have
nonzero CP asymmetries according to the fits.

\TABLE{
\caption{Predicted branching ratios (in units of $10^{-6}$) and CP asymmetries
  for $B_s$ decays in different schemes.  The observables with vanishing
  entries are predicted to be identically zero in our analysis.}
\label{tab:predBs}
\begin{tabular}{ccccc}
\hline\hline
Observable & Scheme 1 & Scheme 2 & Scheme 3 & Scheme 4 \\
\hline
$Br(\pi^{+}\pi^{-})$     &$0$              &$0$              &$0$              &$0$              \\
$Br(\pi^{0}\pi^{0})$     &$0$              &$0$              &$0$              &$0$              \\
$Br(\pi^{+}K^{-})$       &$5.0\pm1.0$              &$5.0\pm1.0$              &$5.0\pm1.0$              &$5.0\pm1.0$              \\
$Br(\pi^{0}K^{0})$       &$1.5\pm0.3$              &$1.5\pm0.3$              &$1.5\pm0.3$              &$1.4\pm0.3$              \\
$Br(K^{+}K^{-})$         &$18.9\pm1.0$             &$18.8\pm1.0$             &$18.8\pm1.0$             &$19.0\pm4.0$             \\
$Br(K^{0}{\bar K}^{0})$         &$20.0\pm1.0$             &$20.2\pm1.0$             &$20.1\pm1.0$             &$20.4\pm4.2$             \\
${\cal A}(\pi^{+}\pi^{-})$     &$0$            &$0$            &$0$            &$0$            \\
${\cal A}(\pi^{0}\pi^{0})$     &$0$            &$0$            &$0$            &$0$            \\
$A_{CP}(\pi^{+}K^{-})$       &$0.32\pm0.07$            &$0.27\pm0.06$            &$0.28\pm0.06$            &$0.26\pm0.06$            \\
${\cal A}(\pi^{0}K_S)$       &$0.47\pm0.15$            &$0.49\pm0.15$            &$0.49\pm0.14$            &$0.50\pm0.14$            \\
${\cal A}(K^{+}K^{-})$         &$-0.08\pm0.02$           &$-0.09\pm0.02$           &$-0.09\pm0.02$           &$-0.09\pm0.02$           \\
${\cal A}(K^{0}{\bar K}^{0})$         &$0$            &$0$            &$0$            &$0$            \\
${\cal S}(\pi^{+}\pi^{-})$      &$0$          &$0$          &$0$          &$0$          \\
${\cal S}(\pi^{0}\pi^{0})$      &$0$          &$0$          &$0$          &$0$          \\
${\cal S}(\pi^{0}K_S)$        &$0.340\pm0.202$          &$0.365\pm0.194$          &$0.359\pm0.193$          &$0.308\pm0.201$          \\
${\cal S}(K^{+}K^{-})$          &$0.147\pm0.022$          &$0.199\pm0.028$          &$0.198\pm0.028$          &$0.211\pm0.035$          \\
${\cal S}(K^{0}{\bar K}^{0})$          &$-0.043\pm0.004$         &$-0.044\pm0.004$         &$-0.044\pm0.004$         &$-0.043\pm0.004$         \\
\hline\hline
\end{tabular}
}

As mentioned above, a good flavor SU(3) symmetry relation has been observed
between $\bar{B}^0 \to K^0 \overline{K}^0$ and $B^- \to \pi^- \bar{K}^0$.  It
is therefore natural to use the $B_s \to K^0 \overline{K}^0$ decay as another
test because it also involves a single $p'$ amplitude.  We predict its
branching ratio to be around $2 \times 10^{-5}$.  Moreover, the time-dependent
CP asymmetries $\cal A$ and $\cal S$ associated with this mode are predicted to
be identically zero and about $-0.044 \pm 0.004$, respectively.  The $B_s \to
K^0 \overline{K}^0$ and $K^+ K^-$ decays have also been discussed in the
literature to study their correlation with the $B_d \to \pi^+ \pi^-$ decay
\cite{Fleischer:2002zv} and their time-dependent CP asymmetries for identifying
new physics \cite{London:2004ej} if they deviate from the SM predictions.

The $B_s \to K^- \pi^+$ and $K^+ K^-$ modes have the same flavor amplitude
decompositions as the ${\bar B}^0 \to \pi^+ \pi^-$ and $\pi^+ K^-$ decays,
respectively.  Therefore, they are predicted to have sizeable CP asymmetries
due to the interference between tree and penguin amplitudes.

The same color-suppressed and penguin amplitudes contribute to both $\bar{B}^0
\to \pi^0 \pi^0$ and $\bar{B}_s \to \pi^0 K^0$ modes.  Neither of them is
dominant in the decay processes.  Therefore, we expect large CP asymmetries in
the latter mode as well.  Moreover, determining the branching ratio of the
latter may provide some insight for the observed large branching ratio of the
former.

\subsection{Fits to all $B \to PP$ modes \label{sec:fitall}}

We further carry out the analysis with the inclusion of modes with $\eta$ and
$\eta'$ in the final state.  In this case, there are totally 34 experimental
observables.  The best fitted parameters are listed in Table~\ref{tab:parmEta}.

\TABLE{
\caption{Fit results of the parameters for all $PP$ modes in Schemes 1 through
  4 defined in the text along with the associated minimal $\chi^2$ values.  The
  amplitudes are given in units of $10^4$ eV.}
\label{tab:parmEta}
\begin{tabular}{ccccc}
\hline\hline
Parameter & Scheme 1 & Scheme 2 & Scheme 3 & Scheme 4 \\
\hline
$\bar{\rho}$                   &$0.089^{+0.031}_{-0.027}$&$0.087^{+0.029}_{-0.026}$&$0.087^{+0.029}_{-0.026}$&$0.096^{+0.029}_{-0.026}$\\       
$\bar{\eta}$                   &$0.377\pm0.027$          &$0.378\pm0.028$          &$0.379\pm0.027$          &$0.370\pm0.027$          \\
$A$                           &$0.809\pm0.012$          &$0.809\pm0.012$          &$0.809\pm0.012$          &$0.809\pm0.012$          \\         
\hline                                                                                                                                  
$|T|$                           &$0.641^{+0.056}_{-0.050}$&$0.642^{+0.056}_{-0.050}$&$0.640^{+0.056}_{-0.049}$&$0.649^{+0.056}_{-0.049}$\\         
$|C|$                           &$0.426\pm0.048$          &$0.418\pm0.048$          &$0.415\pm0.047$          &$0.436\pm0.049$          \\         
$\delta_C$                    &$-72.5\pm7.3$            &$-70.4\pm7.5$            &$-70.0\pm7.3$            &$-68.3\pm7.2$            \\         
$|P|$                           &$0.121\pm0.002$          &$0.121\pm0.002$          &$0.121\pm0.002$          &$0.110\pm0.008$          \\         
$\delta_P$                    &$-17.8\pm3.2$            &$-16.0\pm2.8$            &$-16.4\pm2.8$            &$-15.9\pm2.6$            \\         
$|P_{EW}|$                      &$0.012^{+0.006}_{-0.004}$&$0.011^{+0.005}_{-0.003}$&$0.012^{+0.006}_{-0.004}$&$0.013^{+0.006}_{-0.004}$\\         
$\delta_{P_{EW}}$             &$-58.8^{+39.8}_{-20.6}$  &$-47.7^{+42.9}_{-24.9}$  &$-58.1^{+35.9}_{-19.3}$  &$-57.6^{+32.5}_{-18.2}$  \\         
$|S|$                           &$0.048^{+0.004}_{-0.003}$&$0.047^{+0.004}_{-0.003}$&$0.047^{+0.003}_{-0.003}$&$0.042\pm0.004$          \\         
$\delta_S$                    &$-48.3\pm10.6$           &$-44.8\pm10.2$           &$-44.2\pm9.8$            &$-42.9\pm9.3$            \\         
$\xi$                         &1(fixed)                 &1(fixed)                 &1(fixed)                 &$1.10^{+0.09}_{-0.07}$    \\
\hline
$\delta_{EW}$                 &$0.014\pm0.006$          &$0.013\pm0.005$          &$0.014\pm0.006$          &$0.015\pm0.006$          \\
\hline
$\chi^2/dof$                  &37.4/22                     &34.8/22                     &32.9/22                     &30.6/21                     \\
\hline\hline
\end{tabular}
}

To fit all the $PP$ modes, we have to include at least the flavor singlet
amplitude $S$, whose importance for explaining the large branching ratios of
the $\eta' K$ modes has been noticed and discussed in
Refs.~\cite{DGR,Chiang:2001ir,Fu:2003fy}.  This introduces two more theoretical
parameters $|S|$ and $\delta_S$, the strong phase associated with $S$, than the
fits in Section~\ref{sec:fitsome}.  However, the fitting quality in these
schemes is seen to be much worse than before.  Among the modes with
$\eta^{(\prime)}$ in the final state, $Br(\eta K^-)$, $S_{\eta' K_S}$, and
$A_{CP}(\pi^- \eta')$ have the largest contributions to $\chi^2$.

Comparing the fitting results in Table~\ref{tab:parmEta} with those in
Table~\ref{tab:parafit}, we see that the strong phases suffer from larger
fluctuations among all theoretical parameters.  The values of both $\bar\rho$
and $\bar\eta$ are decreased, but their precisions improved.  This leaves a
smaller region for the $(\bar\rho,\bar\eta)$ vertex, with a $\beta$ consistent
with other observations and a somewhat larger $\gamma$.  The parameters $|T|$
and $|C|$ become slightly larger; but the ratio $|C|/|T| \sim 0.65$ remains
about the same.  The magnitudes of $P$ and $P_{EW}$ (or $\delta_{EW}$) are seen
to be relatively stable in both limited and global fits.  The parameter $\xi$
increases from $1.04$ to $1.10$.

The singlet amplitude has a magnitude about 3 to 4 times $|P_{EW}|$ in our
fits.  Moreover, its strong phase is close to $\delta_{P_{EW}}$ and about
$-30^\circ$ from $P$.  It is this feature that produces interesting
interference patterns among the different modes involving $\eta$ and $\eta'$.

\FIGURE{
\centering
\includegraphics[width=10cm]{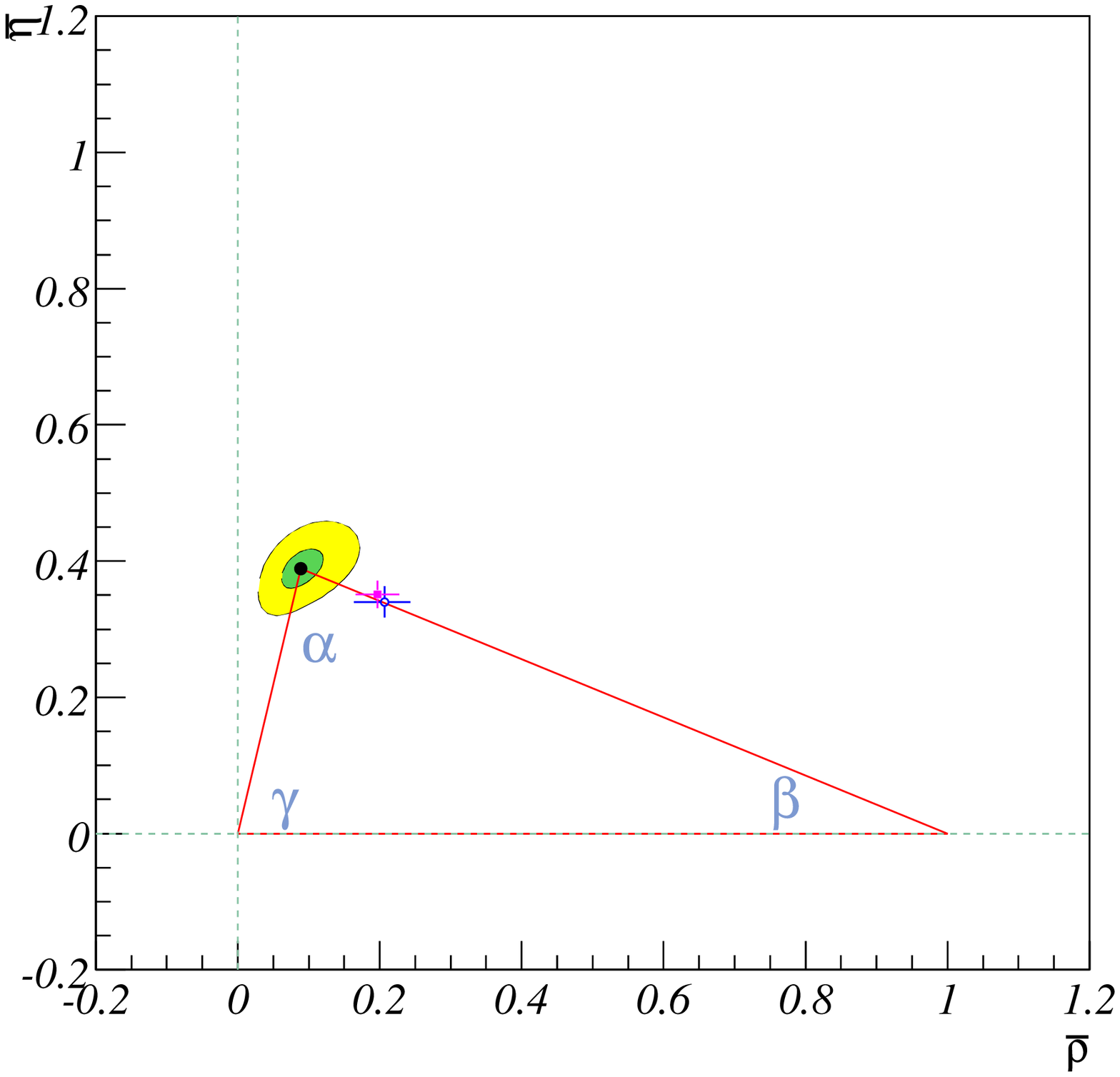}
\caption{Constraints on the $(\bar\rho,\bar\eta)$ vertex using all the $PP$
  mode data in Scheme~3 defined in the text.  Contours correspond to 1 $\sigma$
  and $95\%$ CL, respectively.  The crosses refer to the 1 $\sigma$ range given
  by the latest CKMfitter (open circle) and UTfit (filled square) results using
  other methods \cite{CKMfitter,UTfit} as a comparison.}
\label{fig:rhoeta-all}
}

Scheme~3 in this case gives the following results for the weak phases $\alpha$,
$\beta$, and $\gamma$:
\begin{eqnarray}
  &
  \alpha = \left( 80 \pm 6 \right)^\circ ~, \quad\mbox{or}\quad
  69^\circ \le \alpha \le 92^\circ \quad \mbox{(95\% CL)} ~;
  &
  \nonumber \\
  &
  \beta = \left( 23 \pm 2 \right)^\circ ~, \quad\mbox{or}\quad
  20^\circ \le \beta \le 27^\circ \quad \mbox{(95\% CL)} ~;
  &
  \\
  &
  \gamma = \left( 77 \pm 4 \right)^\circ ~, \quad\mbox{or}\quad
  69^\circ \le \gamma \le 84^\circ \quad \mbox{(95\% CL)} ~.
  &
  \nonumber
\end{eqnarray}

The predictions of the $B_{u,d}$ decay observables are given in
Table~\ref{tab:brcpEta}.  Vanishing observables in our approach are omitted to
avoid an oversized table.  The following observables deviate the most from the
current data: ${\cal S}(\pi^0 K_S)$, $Br(\pi^- \pi^0)$, $Br(K^- \eta)$,
$Br(\pi^0 \pi^0)$, ${\cal S}(\eta' K_S)$, and $A_{CP}(\pi^-\eta)$ (listed in
the order of their contributions to $\chi^2$).

As in the limited fits in Section~\ref{sec:fitsome}, the global fits also
prefer sizeable CP asymmetries for the $\pi^0 \pi^0$ and $\pi^+ \pi^-$ modes.
The branching ratios of the yet measured $\eta\eta$, $\eta\eta'$, and
$\eta'\eta'$ are all consistent with the current upper bounds.  Their
corresponding direct CP asymmetries are predicted to be large.  However,
measuring them will require more work.

The current branching ratios of $\pi^0 \eta$ and $\pi^0 \eta'$ has a factor of
2 difference in the central values, though the errors are still large.  Our
results, on the other hand, show that they are equal to each other in all
schemes.  ${\cal A}(\eta K_S)$ has not been observed, but is 2 $\sigma$ away
from zero in our analysis.

\TABLE{
\caption{Predicted branching ratios (in units of $10^{-6}$) and CP asymmetries
  for all the $PP$ modes of $B^{0,+}$ decays in different schemes.  Vanishing
  observables in our approach are omitted.}
\label{tab:brcpEta}
\begin{tabular}{ccccc}
\hline\hline
Observable & Scheme 1 & Scheme 2 & Scheme 3 & Scheme 4  \\
\hline
$Br(\pi^+\pi^-)$         &$5.3\pm1.0$              &$5.3\pm1.0$              &$5.3\pm1.0$              &$5.3\pm1.0$              \\
$Br(\pi^0\pi^0)$         &$1.7\pm0.3$              &$1.7\pm0.3$              &$1.7\pm0.3$              &$1.6\pm0.3$              \\
$Br(\pi^-\pi^0)$         &$4.8\pm1.0$              &$4.9\pm1.0$              &$4.9\pm1.0$              &$5.1\pm1.1$              \\
$Br(\pi^+K^-)$           &$20.3\pm1.0$             &$20.2\pm1.0$             &$20.2\pm1.0$             &$20.4\pm4.3$             \\
$Br(\pi^0\bar{K}^0)$     &$9.6\pm1.0$              &$9.6\pm0.9$              &$9.6\pm1.0$              &$9.8\pm2.3$              \\
$Br(\pi^-\bar{K}^0)$     &$22.6\pm1.1$             &$22.7\pm1.1$             &$22.7\pm1.1$             &$23.1\pm4.8$             \\
$Br(\pi^0K^-)$           &$12.3\pm1.2$             &$12.2\pm1.1$             &$12.3\pm1.2$             &$12.5\pm2.7$             \\
$Br(K^0\bar{K}^0)$       &$1.1\pm0.1$              &$1.1\pm0.1$              &$1.1\pm0.1$              &$0.9\pm0.1$              \\
$Br(K^-\bar{K}^0)$       &$1.2\pm0.1$              &$1.2\pm0.1$              &$1.2\pm0.1$              &$1.0\pm0.2$              \\
$Br(\pi^0\eta)$          &$1.0\pm0.1$              &$1.0\pm0.1$              &$1.0\pm0.1$              &$0.8\pm0.1$              \\
$Br(\pi^0\eta')$         &$1.0\pm0.1$              &$1.0\pm0.1$              &$1.0\pm0.1$              &$0.8\pm0.1$              \\
$Br(\pi^-\eta)$          &$4.6\pm0.6$              &$4.6\pm0.6$              &$4.6\pm0.6$              &$4.6\pm0.7$              \\
$Br(\pi^-\eta')$         &$3.2\pm0.3$              &$3.2\pm0.3$              &$3.2\pm0.3$              &$3.0\pm0.4$              \\
$Br(\bar{K}^0\eta)$      &$1.4\pm0.2$              &$1.3\pm0.2$              &$1.4\pm0.2$              &$1.4\pm0.3$              \\
$Br(\bar{K}^0\eta')$     &$65.3\pm5.2$             &$65.7\pm5.0$             &$65.5\pm4.8$             &$66.4\pm13.0$            \\
$Br(K^-\eta)$            &$1.5\pm0.3$              &$1.5\pm0.2$              &$1.5\pm0.3$              &$1.5\pm0.4$              \\
$Br(K^-\eta')$           &$69.2\pm5.5$             &$69.5\pm5.3$             &$69.3\pm5.1$             &$70.1\pm13.8$            \\
$Br(\eta\eta)$           &$0.8\pm0.1$              &$0.8\pm0.1$              &$0.8\pm0.1$              &$0.8\pm0.1$              \\
$Br(\eta'\eta')$         &$0.4\pm0.0$              &$0.4\pm0.0$              &$0.4\pm0.0$              &$0.4\pm0.0$              \\
$Br(\eta\eta')$          &$1.2\pm0.1$              &$1.2\pm0.1$              &$1.2\pm0.1$              &$1.1\pm0.1$              \\
${\cal A}(\pi^+\pi^-)$         &$0.27\pm0.06$            &$0.24\pm0.05$            &$0.25\pm0.05$            &$0.22\pm0.04$            \\
${\cal A}(\pi^0\pi^0)$         &$0.71\pm0.10$            &$0.70\pm0.10$            &$0.70\pm0.10$            &$0.67\pm0.09$            \\
$A_{CP}(\pi^-\pi^0)$         &$0.03\pm0.03$            &$0.02\pm0.03$            &$0.03\pm0.03$            &$0.04\pm0.03$            \\
$A_{CP}(\pi^+K^-)$           &$-0.07\pm0.02$           &$-0.08\pm0.02$           &$-0.08\pm0.02$           &$-0.08\pm0.02$           \\
${\cal A}(\pi^0 K_S)$     &$-0.13\pm0.02$           &$-0.12\pm0.02$           &$-0.15\pm0.03$           &$-0.17\pm0.03$           \\
$A_{CP}(\pi^-\eta)$          &$-0.09\pm0.10$           &$-0.11\pm0.09$           &$-0.10\pm0.09$           &$-0.10\pm0.09$           \\
$A_{CP}(\pi^-\eta')$         &$0.06\pm0.12$            &$0.04\pm0.12$            &$0.04\pm0.12$            &$0.02\pm0.11$            \\
${\cal A}(\eta K_S)$      &$0.13\pm0.07$            &$0.14\pm0.07$            &$0.16\pm0.08$            &$0.18\pm0.09$            \\
${\cal A}(\eta' K_S)$     &$0.02\pm0.00$            &$0.02\pm0.00$            &$0.03\pm0.01$            &$0.03\pm0.01$            \\
$A_{CP}(K^-\eta)$            &$-0.25\pm0.12$           &$-0.29\pm0.13$           &$-0.27\pm0.14$           &$-0.29\pm0.15$           \\
${\cal A}(\eta\eta)$           &$-0.77\pm0.11$           &$-0.78\pm0.11$           &$-0.76\pm0.11$           &$-0.73\pm0.11$           \\
${\cal A}(\eta'\eta')$         &$-0.55\pm0.13$           &$-0.55\pm0.13$           &$-0.55\pm0.13$           &$-0.58\pm0.13$           \\
${\cal A}(\eta\eta')$          &$-0.65\pm0.13$           &$-0.66\pm0.13$           &$-0.66\pm0.12$           &$-0.66\pm0.12$           \\
${\cal S}(\pi^+\pi^-)$          &$-0.533\pm0.135$         &$-0.533\pm0.135$         &$-0.532\pm0.134$         &$-0.513\pm0.138$         \\
${\cal S}(\pi^0\pi^0)$          &$0.634\pm0.116$          &$0.655\pm0.111$          &$0.649\pm0.111$          &$0.614\pm0.118$          \\
${\cal S}(\pi^0 K_S)$      &$0.780\pm0.041$          &$0.781\pm0.041$
&$0.789\pm0.041$          &$0.791\pm0.041$         \\
${\cal S}(K^0\bar{K}^0)$        &$-0.001\pm0.031$         &$-0.001\pm0.031$
&$-0.001\pm0.030$         &$-0.000\pm0.017$         \\
${\cal S}(\eta K_S)$ &$0.5\pm0.06$ &$0.5\pm0.05$ &$0.45\pm0.06$ &$0.39\pm0.07$\\
${\cal S}(\eta' K_S)$ &$0.72\pm0.04$ &$0.72\pm0.04$ &$0.72\pm0.04$ &$0.71\pm0.04$\\
\hline\hline
\end{tabular}
}

The predictions for the $B_s$ modes are given in Table~\ref{tab:brcpBsEta}.
Many of the discussions regarding the modes with $\pi$ and $K$ mesons in the
final state in Section~\ref{sec:fitsome} can be applied here.  Thus, we only
concentrate on the modes with $\eta$ and/or $\eta'$ in the final state.

As in the cases of ${\bar B}^0 \to \eta' K_S$ and $B^- \to K^- \eta'$, the
constructive interference between $p'$ and $s'$ makes the ${\bar B}_s^0 \to
\eta'\eta'$ the one with the largest branching ratio, $\sim 50 \times 10^{-6}$,
among all.  The same effect is seen in the ${\bar B}_s^0 \to \eta \eta'$ decay
as well.  In contrast, a destructive effect occurs in the ${\bar B}_s^0 \to
\eta \eta$ decay, so that its branching ratio is only $\sim 2 \times 10^{-6}$.

The ${\bar B_s}^0 \to \eta' K_S$ decay is another place where the constructive
interference between the QCD penguin and singlet penguin amplitudes plays an
important role.  Although small in magnitude for $\Delta S = 0$ transitions,
they can interfere with the color-suppressed amplitude to give potentially
observable time-dependent CP asymmetries, both predicted at $\sim 5 \sigma$
level.

\TABLE{
\caption{Predicted branching ratios (in units of $10^{-6}$) and CP asymmetries
  for all the $PP$ modes of $B_s$ decays in different schemes.  Observables
  with vanishing entries are omitted.}
\label{tab:brcpBsEta}
\begin{tabular}{ccccc}
\hline\hline
Observable & Scheme 1 & Scheme 2 & Scheme 3 & Scheme 4 \\
\hline
$Br(\pi^{+}\pi^{-})$     &$0$              &$0$              &$0$              &$0$              \\
$Br(\pi^{0}\pi^{0})$     &$0$              &$0$              &$0$              &$0$              \\
$Br(\pi^{+}K^{-})$       &$5.0\pm0.9$              &$5.0\pm0.9$              &$5.0\pm0.9$              &$5.0\pm0.9$              \\
$Br(\pi^{0}K^{0})$       &$1.6\pm0.3$              &$1.6\pm0.3$              &$1.6\pm0.3$              &$1.5\pm0.3$              \\
$Br(K^{+}K^{-})$         &$18.9\pm1.0$             &$18.9\pm1.0$             &$18.9\pm1.0$             &$19.1\pm4.0$             \\
$Br(K^{0}K^{0})$         &$19.7\pm1.0$             &$19.8\pm1.0$             &$19.8\pm1.0$             &$20.2\pm4.2$             \\
$Br(\pi^0\eta)$          &$0$              &$0$              &$0.1\pm0.0$              &$0.1\pm0.0$              \\
$Br(\pi^0\eta')$         &$0.1\pm0.0$              &$0.1\pm0.0$              &$0.1\pm0.0$              &$0.1\pm0.1$              \\
$Br(\bar{K}^0\eta)$      &$0.7\pm0.2$              &$0.7\pm0.2$              &$0.7\pm0.2$              &$0.7\pm0.2$              \\
$Br(\bar{K}^0\eta')$     &$3.3\pm0.3$              &$3.4\pm0.3$              &$3.4\pm0.3$              &$2.8\pm0.3$              \\
$Br(\eta\eta)$           &$2.0\pm0.4$              &$2.0\pm0.4$              &$2.0\pm0.4$              &$2.0\pm0.6$              \\
$Br(\eta'\eta')$         &$48.3\pm4.4$             &$48.6\pm4.3$             &$48.3\pm4.1$             &$48.9\pm9.8$             \\
$Br(\eta\eta')$          &$22.4\pm1.5$             &$22.6\pm1.4$             &$22.5\pm1.4$             &$22.9\pm4.7$             \\
${\cal A}(\pi^{+}\pi^{-})$     &$0$            &$0$            &$0$            &$0$            \\
${\cal A}(\pi^{0}\pi^{0})$     &$0$            &$0$            &$0$            &$0$            \\
$A_{CP}(\pi^{+}K^{-})$       &$0.27\pm0.06$            &$0.24\pm0.05$            &$0.25\pm0.05$            &$0.22\pm0.04$            \\
${\cal A}(\pi^{0}K_S)$       &$0.71\pm0.10$            &$0.70\pm0.10$            &$0.70\pm0.10$            &$0.67\pm0.09$            \\
${\cal A}(K^{+}K^{-})$         &$-0.07\pm0.02$           &$-0.08\pm0.02$           &$-0.08\pm0.02$           &$-0.08\pm0.02$           \\
${\cal A}(K^{0}K^{0})$         &$0$            &$0$            &$0$            &$0$            \\
${\cal A}(\pi^0\eta)$          &$0.20\pm0.47$            &$0.32\pm0.48$            &$0.19\pm0.46$            &$0.18\pm0.45$            \\
${\cal A}(\pi^0\eta')$         &$0.20\pm0.47$            &$0.32\pm0.48$            &$0.19\pm0.46$            &$0.18\pm0.45$            \\
${\cal A}(\eta K_S)$      &$-0.24\pm0.13$           &$-0.27\pm0.13$           &$-0.26\pm0.12$           &$-0.22\pm0.11$           \\
${\cal A}(\eta' K_S)$     &$-0.42\pm0.08$           &$-0.41\pm0.08$           &$-0.41\pm0.08$           &$-0.45\pm0.08$           \\
${\cal A}(\eta\eta)$           &$-0.20\pm0.03$           &$-0.20\pm0.03$           &$-0.24\pm0.04$           &$-0.27\pm0.05$           \\
${\cal A}(\eta'\eta')$         &$0.03\pm0.01$            &$0.03\pm0.01$            &$0.03\pm0.01$            &$0.04\pm0.01$            \\
${\cal A}(\eta\eta')$          &$-0.02\pm0.00$           &$-0.02\pm0.00$           &$-0.03\pm0.01$           &$-0.03\pm0.01$           \\
${\cal S}(\pi^{+}\pi^{-})$      &$0$          &$0$          &$0$          &$0$          \\
${\cal S}(\pi^{0}\pi^{0})$      &$0$          &$0$          &$0$          &$0$          \\
${\cal S}(\pi^{0}K_S)$        &$0.282\pm0.158$          &$0.318\pm0.153$          &$0.311\pm0.153$          &$0.185\pm0.167$          \\
${\cal S}(K^{+}K^{-})$          &$0.167\pm0.024$          &$0.217\pm0.030$          &$0.216\pm0.030$          &$0.244\pm0.037$          \\
${\cal S}(K^{0}K^{0})$          &$-0.041\pm0.004$         &$-0.041\pm0.004$
&$-0.041\pm0.004$         &$-0.040\pm0.003$         \\
${\cal S}(\eta K_S)$ &$0.26\pm0.17$ &$0.24\pm0.17$ &$0.26\pm0.17$ &$0.26\pm0.16$\\
${\cal S}(\eta' K_S)$ &$0.40\pm0.08$ &$0.39\pm0.08$ &$0.39\pm0.08$ &$0.45
\pm0.08$\\
\hline\hline
\end{tabular}
}

\subsection{Fits with a new physics amplitude \label{sec:fitNP}}

In expectation of possible new physics contributions to the $K \pi$ decays to
account for the observed branching ratio and CP violation pattern
\cite{Yoshikawa,Buras:2003yc,Barger:2004hn,Baek:2006ti,Gronau:2006xu}, we try
in Scheme~3 fits with a new amplitude added to these decays.  More explicitly,
a new amplitude $N = |N| \exp{\left[ i (\phi_N + \delta_N) \right]}$ is
included in the $B \to \pi^0 K^-$ and $\pi^0 \bar{K}^0$ decays in such a way
that effectively,
\begin{eqnarray}
  c' \to Y_{sb}^u \xi_c C - \left( Y_{sb}^u + Y_{sb}^c \right) P_{EW} + N ~.
\end{eqnarray}
This introduces three more parameters ($|N|$, $\phi_N$, and $\delta_N$) into
the fits.  Here we assume that $P_{EW}$ is fixed relative to $T+C$ through the
SM relation.  $\chi^2_{\rm min}$ is found to decrease from $16.4$ to $4.3$ in
the limited fit with only $\pi$, $K$ mesons in the final state.  The new
physics parameters are found to be
\begin{eqnarray}
\label{eq:NPpara}
|N| = 18^{+3}_{-4} \; \mbox{eV} ~, \quad 
\phi_N=(92\pm 4)^\circ ~, \quad \mbox{and} \quad
\delta_N=(-14\pm 5)^\circ ~.
\end{eqnarray}
The best fitted CKM parameters $\bar\rho$, $\bar\eta$ and $A$ remain almost
unchanged.  The best fitted tree-type amplitudes have $|T| =
(0.55^{+0.05}_{-0.04}) \times 10^4$~eV, $|C| = (0.32^{+0.05}_{-0.04}) \times
10^4$~eV and $\delta_C = (-39^{+16}_{-13})^\circ$.  The penguin amplitude $|P|$
is unchanged.

Since both $C$ and $P_{EW}$ have the same flavor topology, they always appear
in pairs in $c$ or $c'$ for any physical decay process.  Therefore, it seems
difficult to determine whether the new amplitude is associated with one or the
other \cite{Imbeault:2006nx}.  Our results have $|N| / |V_{cb}V_{cs}| \simeq
0.04\times 10^4\mbox{eV}$ and $|N| / |V_{ub}V_{us}| \simeq 2.2\times
10^4\mbox{eV}$, showing that the new amplitude is unexpectedly large.  It is
about five times bigger than $|P_{EW}|$ or $|C|$.  Since we assume that it only
enters $c'$ in the $K \pi$ modes instead of $c$ in the $\pi \pi$ modes, this
result suggests that it behaves more like the electroweak penguin amplitude
than the color-suppressed amplitude, for the former plays a much less important
role than the latter in the strangeness-conserving modes.  Although the latest
data indicate only a mild deviation of $\pi K$ branching ratios from SM
estimates, the large difference between  $A_{CP}(\pi^+K^-)$ and $A_{CP}(\pi^0
K^-)$ is unexpected, which require a much larger $|C'|/|T'|$ than $|C|/|T|$ in
$\pi\pi$ modes \cite{Wu:2006yj}.  Within the framework of SU(3) symmetry, a
large new physics contribution is still possible.

The above conclusion may look contradictory to what we have found in
Section~\ref{sec:fitsome},  where $|P_{EW}|$ is preferred by data to fall
within the SM expectation, meaning that varying its value would not improve the
fitting quality.  But this is only because in the previous fit, the weak phase
of $P_{EW}$ is fixed to the SM value.  In the analysis of this section, the
electroweak penguin-like new amplitude $N$ is allowed to have its own weak and
strong phases.

It should be emphasized that $N$ is not added to modes with the contributing
amplitude $c'$ in the global fits other than the $K \pi$ decays.  It does not
improve the minimal $\chi^2$ much to do so, for there is a pull between the $K
\pi$ and $K \eta^{(\prime)}$ data such that $|N|$ is about a factor of 5
smaller than that quoted above.  Therefore, it remains to be understood why the
new amplitude does not help when modes with $\eta$ and $\eta'$ are taken into
consideration as it should if flavor SU(3) is respected.  The solution to this
question may rely on more precisely determined branching ratios of $\eta
K^{0,-}$.

\section{Summary \label{sec:summary}}

In this paper, we perform $\chi^2$ fits to the branching ratios and CP
asymmetries of both limited and entire sets of the rare $B \to PP$ decays.  We
consider the primary contributing flavor amplitudes $T$, $C$, $P$, and
$P_{EW}$, each of which is associated with a distinct strong phase.  The
analysis is based upon flavor SU(3) symmetry.  We also include the $f_K /
f_\pi$ ratio and an additional SU(3) breaking factor $\xi$ to test the
stability of our fits.

One major result is the extraction of the vertex $(\bar\rho,\bar\eta)$ and thus
the weak phases of the CKM unitarity triangle.  This is complementary to other
methods.  The values of $\beta$ and $\gamma$ obtained from our fits for modes
without $\eta^{(\prime)}$ in the final state are generally larger than but
still consistent within errors with those given by the overall fits of the
CKMfitter and UTfit groups to other observables.  Our fits to all the $PP$
modes, however, result in a $\beta$ similar to that given by both the CKMfitter
and the UTfit groups but a larger $\gamma$.

The current $PP$ data favor a large $C$ with a strong phase of about
$-60^\circ$ with respect to $T$.  This is seen to be required by the large
branching ratio of $\pi^0 \pi^0$ and the fact that $A_{CP}(\pi^0 K^-)$ is
different from $A_{CP}(\pi^+ K^-)$.  On the other hand, the size of electroweak
penguin amplitude $P_{EW}$ is found to be consistent with the SM expectation.

We also comment on the possibility of a new physics contribution to $K \pi$
decays.  Our fitting analysis in this case prefers an electroweak penguin-like
amplitude with sizeable magnitude and nontrivial weak phase given in
Eq.~(\ref{eq:NPpara}).  However, this amplitude does not respect flavor SU(3)
symmetry.

Using the parameters extracted from fitting, we make predictions for all the
rare $B \to PP$ processes.  Moreover, we extend our predictions to the
observables in $B_s$ decays based on flavor SU(3) symmetry, whose experimental
data will become available for comparison in the next couple of years from
Tevatron Run II, LHCb and upgraded Belle experiments.

\section*{Acknowledgments}

C.-W.~C. would like to thank the hospitality of the KEK Theory Group during his
visit where this work was initiated and that of the National Center for
Theoretical Sciences in Hsinchu.  He also thanks N.~Deshpande for helpful
discussions and the hospitality of the Institute of Theoretical Science at
Univ. of Oregon during his visit.  This research was supported in part by the
National Science Council of Taiwan, R.O.C.\ under Grant No.\ NSC
94-2112-M-008-023-.  Y.-F.~Z. is supported by JSPS foundation.


\end{document}